\documentclass[aps,prl,preprint,superscriptaddress]{revtex4-2}
\usepackage{graphicx} 
\usepackage{amsmath}
\usepackage{xcolor}
\usepackage{hyperref}
\begin{document}

\date{\today}

\title{Edge polaritons at metal-insulator boundaries in a phase separated correlated oxide}

\author{Weiwei Luo}
\thanks{These authors contributed equally to this work.}
\affiliation{The Key Laboratory of Weak-Light Nonlinear Photonics, Ministry of Education, School of Physics and TEDA Applied Physics Institute, Nankai University, Tianjin 300457, China}
\affiliation{Department of Quantum Matter Physics, University of Geneva, Geneva 1211, Switzerland}

\author{Adrien Bercher}
\thanks{These authors contributed equally to this work.}
\affiliation{Department of Quantum Matter Physics, University of Geneva, Geneva 1211, Switzerland}

\author{Claribel Dominguez}
\affiliation{Department of Quantum Matter Physics, University of Geneva, Geneva 1211, Switzerland}

\author{Javier del Valle}
\affiliation{Department of Physics, University of Oviedo, C/ Federico García Lorca 18, 33007 Oviedo, Spain}
\affiliation{Center of Research on Nanomaterials and Nanotechnology, CINN (CSIC-Universidad de Oviedo), El Entrego 33940, Spain}

\author{J\'{e}r\'{e}mie Teyssier}
\affiliation{Department of Quantum Matter Physics, University of Geneva, Geneva 1211, Switzerland}

\author{Javier Taboada-Gutiérrez}
\affiliation{Department of Quantum Matter Physics, University of Geneva, Geneva 1211, Switzerland}

\author{Alexey B. Kuzmenko}
\email{Alexey.Kuzmenko@unige.ch}
\affiliation{Department of Quantum Matter Physics, University of Geneva, Geneva 1211, Switzerland}

\begin{abstract}

Correlated transition metal oxides, such as cuprates, nickelates, and manganites, are typically considered "bad metals", where high electromagnetic losses suppress the conventional plasmonic effects observed in noble metals, 2D electron gases, and graphene. Nevertheless, using mid-infrared near-field optical nanoscopy, we demonstrate the emergence of strongly confined and long-propagating edge polaritons (EPs) of mixed phonon-plasmon nature at the boundaries between conducting and insulating regions in thin NdNiO$_{3}$ films, fingerprinted as a pronounced peak of the near-field signal phase. Our simulations reveal that the electromagnetic nature of the EPs depends significantly on the edge smoothness, being caused by a one-dimensional optical edge state (ES) at abrupt edges while being governed by the epsilon-near-zero (ENZ) absorption in the case of broad boundaries. Our findings highlight the critical role of nonlocal plasmonic effects in near-field imaging of phase-separated correlated oxides and open new avenues for infrared plasmonics in this family of materials.
\end{abstract}

\maketitle

\noindent \textbf{\large Introduction}

\noindent Metal-insulator transitions (MITs) driven by temperature, strain and other control parameters present a fascinating and practically important phenomenon in solid state physics \cite{ImadaRMP98}. Materials undergoing a first-order MIT often exhibit phase-separation, spontaneously forming metal-insulator (MI) boundaries \cite{QazilbashScience07,PostNP18}. These structures are potentially interesting for nanophotonics, as the sign change of the dielectric permittivity across them gives rise to the emergence of polaritons (hybrid light-matter electromagnetic modes) confined to the interface \cite{MaierBook07,NovotnyBook12}. Moreover, the boundaries can be two-dimensional (in bulk samples) or quasi-one-dimensional (in thin films), which is important since the dimensionality is a key parameter in the physics of polaritons. Indeed, a two-dimensional (2D) metal-insulator boundary between two bulk media hosts surface plasmon polaritons (SPPs, polaritons resulting from the coupling of light with the electron cloud of a metal), which are widely exploited for biosensing and other applications \cite{EnglebienneBook03,SchasfoortBook17}. These polaritons propagate along the boundary and are fundamentally distinct from the bulk three-dimensional polaritons \cite{FetterPRB86}. Likewise, in the case of ultra-thin metallic films, the one-dimensional boundary can host edge plasmon-polaritons (EPPs), which differ significantly from the SPPs. EPPs have been observed spectroscopically in metallic thin films \cite{CamposACSPh17,BellidoACSPh16}, two dimensional electron gases (2DEGs)\cite{MastPRL85}, graphene \cite{NikitinNPh16} and topological insulators \cite{ChenNC22}. Additionally, edge-confined phonon polaritons have been reported in polar van der Waals materials \cite{ZhaoOE16,LiNL17,DaiAM18,ChenAOM22,SunNN24}. 

Observing edge polaritons at MI boundaries in spontaneously phase-separated systems may seem improbable, because the strongly correlated electronic systems where such transitions typically occur have intrinsic electromagnetic losses much higher than those in conventional plasmonic materials. Nonetheless, using scattering-type scanning near-field optical microscopy (s-SNOM), we observe a pronounced near-field phase peak at the MI boundaries in ultra-thin films (10 nm and 40 nm in thickness) of NdNiO$_3$ (NNO), where the transition is induced by either temperature changes or local Joule heating from an electric current. Based on electromagnetic simulations, we attribute this peak to the presence of the edge-confined mixed plasmon-phonon polaritons, or shortly by edge polaritons, with the electromagnetic nature being dependent on the boundary smoothness. 

\hspace{2pt}

\noindent \textbf{\large DC transport and infrared spectra}

\begin{figure*}[htb]
\centerline{\includegraphics[width=18cm]{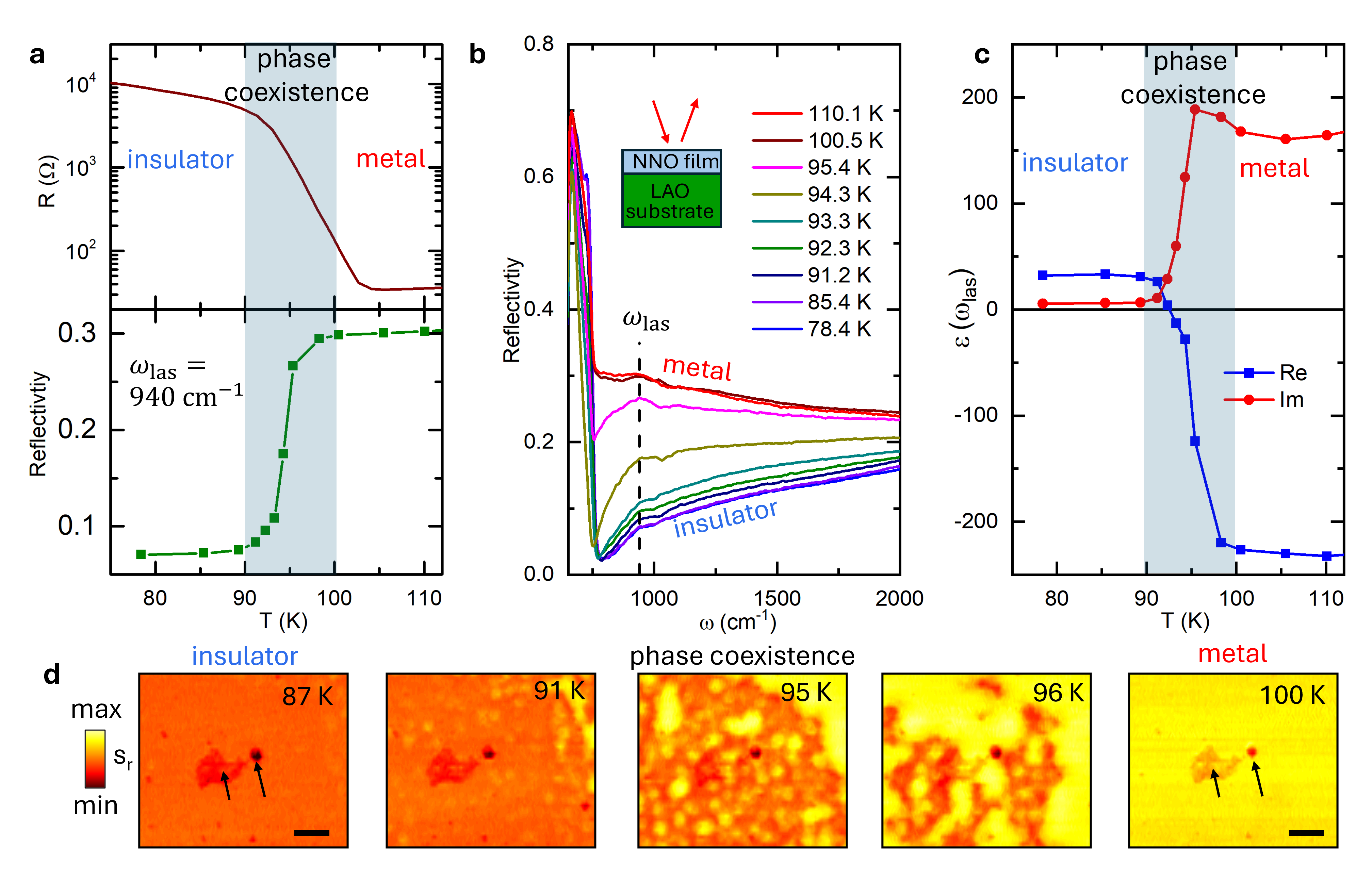}}
\caption{\textbf{Evolution of the electrical and infrared properties of a 10 nm thick NdNiO$_{3}$ film on LaAlO$_{3}$ across the MIT.} \textbf{a}, Temperature dependence of the sample resistance (top panel) and the far-field reflectivity at the frequency $\omega_{\text{las}}=940\,\text{cm}^{-1}$ (bottom panel). \textbf{b}, Near-normal incidence reflectivity spectra measured at selected temperatures. \textbf{c}, Real and imaginary parts of the dielectric function of the film at the same frequency, extracted from a Kramers-Kronig consistent analysis of the reflectivity spectra. \textbf{d}, s-SNOM images at different temperatures of the ratio between the amplitudes of the third- and the second harmonics $s_{r}=s_{3}/s_{2}$. An impurity phase and a dust particle are shown with arrows. The scale bar is 1~$\mu$m. In both far- and near-field measurements, a gold-covered region of the sample was used as reference. The gray boxes in \textbf{a} and \textbf{c} indicate the thermal regime with phase separation.}
\label{Fig1}
\end{figure*}

\noindent Fig. \ref{Fig1} presents the effect of the MIT on spatially averaged transport and infrared properties in a 10 nm (25 unit cell) thin NNO film having been epitaxially grown on a (100) oriented substrate of LaAlO$_3$ (LAO) (see Methods). At cooling down, the film resistance shows a sharp metal-to-insulator transition with a width of about 10 K, centered at around $T_{\text{MIT}}$ = 95 K  (top part of Fig. \ref{Fig1}a). This transition is manifestly first order \cite{TorrancePRB92,MedardeJPCM97,CatalanoRPP18}, with a thermal hysteresis of about 25 K, but for the purpose of this paper we discuss only the cooling stage. 


Infrared reflectivity spectra (see Methods) obtained at various temperatures across the MIT (Fig. \ref{Fig1}b) show that the electromagnetic response below 750 cm$^{-1}$ is dominated by the optical phonons in LAO, while at higher frequencies it is mostly determined by the properties of the film. The temperature dependence of the reflectivity at $\omega_{\text{las}}$ = 940 cm$^{-1}$ (116 meV, 10.64 $\mu$m), which corresponds to one of the laser wavelengths used in the s-SNOM (bottom part of Fig. \ref{Fig1}a) corroborates the sharpness of the transition seen in the transport measurements. By applying a Kramers-Kronig constrained analysis to the broadband reflectivity spectra (see Methods), we obtain the real and the imaginary parts of the complex dielectric function $\varepsilon$ of the film (Supplementary Fig. S1). In Fig. \ref{Fig1}c we present the thermal evolution of these quantities at $\omega_{\text{las}}$. One can see that above the transition, $\text{Re}(\varepsilon) \approx -230$ and $\text{Im}(\varepsilon)\approx 160$, indicating a metallic state with a high inductance but also high electromagnetic losses. Below $T_{\text{MIT}}$,  $\text{Re}(\varepsilon) \approx 30$ and $\text{Im}(\varepsilon)\approx 5$, reflecting an insulating state with a high refractive index and moderate losses. 

Several previous studies have shown that thin NNO films undergo a phase separation during the transition \cite{MattoniNC16,PostNP18,LeeNL19,LiNC19}. Indeed, temperature-dependent nanoscale optical imaging on our sample clearly reveals the presence of an inhomogeneous state close to the MIT (Fig. \ref{Fig1}d). Briefly, in the s-SNOM experiment, a sharp metal coated atomic-force microscopy (AFM) tip is placed closely to the surface while being illuminated with an external laser source. Via the antenna effect, the tip concentrates electromagnetic field at the very apex, which interacts with the sample in the near-field, \textit{i. e.} on a scale much smaller than the laser wavelength \cite{KnollNature99,TaubnerNL04,ChenAM19} setting an unprecedented optical resolution limited only by the tip radius (20-40 nm). The radiation back-scattered by the tip contains information about the sample's local optical properties and, via interferometry, both amplitude and phase of the near-field complex-valued optical signal are measured. The AFM operates in tapping mode, and the signal is demodulated at higher harmonics of the tapping frequency, $\tilde{s}_{n}=s_{n}e^{i\theta_{n}}$, allowing separation of the near-field signal from the far-field background. We further normalize the signal using the ratio between the amplitudes of the third and the second s-SNOM harmonics, $\tilde{s}_{r}=\tilde{s}_{3}/\tilde{s}_{2}=s_{r}e^{i\theta_{r}}$, which removes unwanted signals related to slow spatial variation of the tip illumination \cite{MesterNanophotonics22}. We can see from the s-SNOM amplitude maps of $s_{r}$ (Fig. \ref{Fig1}d) that the metallic (high signal) and insulating (low signal) phases coexist between 90 and 100 K, perfectly matching the transition temperature interval measured by DC transport and far-field spectroscopy (Fig. \ref{Fig1}a). 

Due to the phase separation, the dielectric functions extracted in the intermediate temperature range should be considered as  mean effective-medium values not necessarily corresponding to a pure metallic or a pure insulator states. Therefore, in the following analysis we use the extracted dielectric functions of the film in its homogeneous states just below and just above the transition as the intrinsic mono-phase values. Specifically, at the s-SNOM laser frequency we have $\varepsilon_{\text{met}}= -230 + 160 i$ and $\varepsilon_{\text{ins}}= 30 + 5 i$ for the metallic and insulating states respectively.

\hspace{2pt}

\noindent \textbf{\large Correlative amplitude-phase analysis of the experimental s-SNOM maps}

\begin{figure*}[htb]
\centerline{\includegraphics[width=18cm]{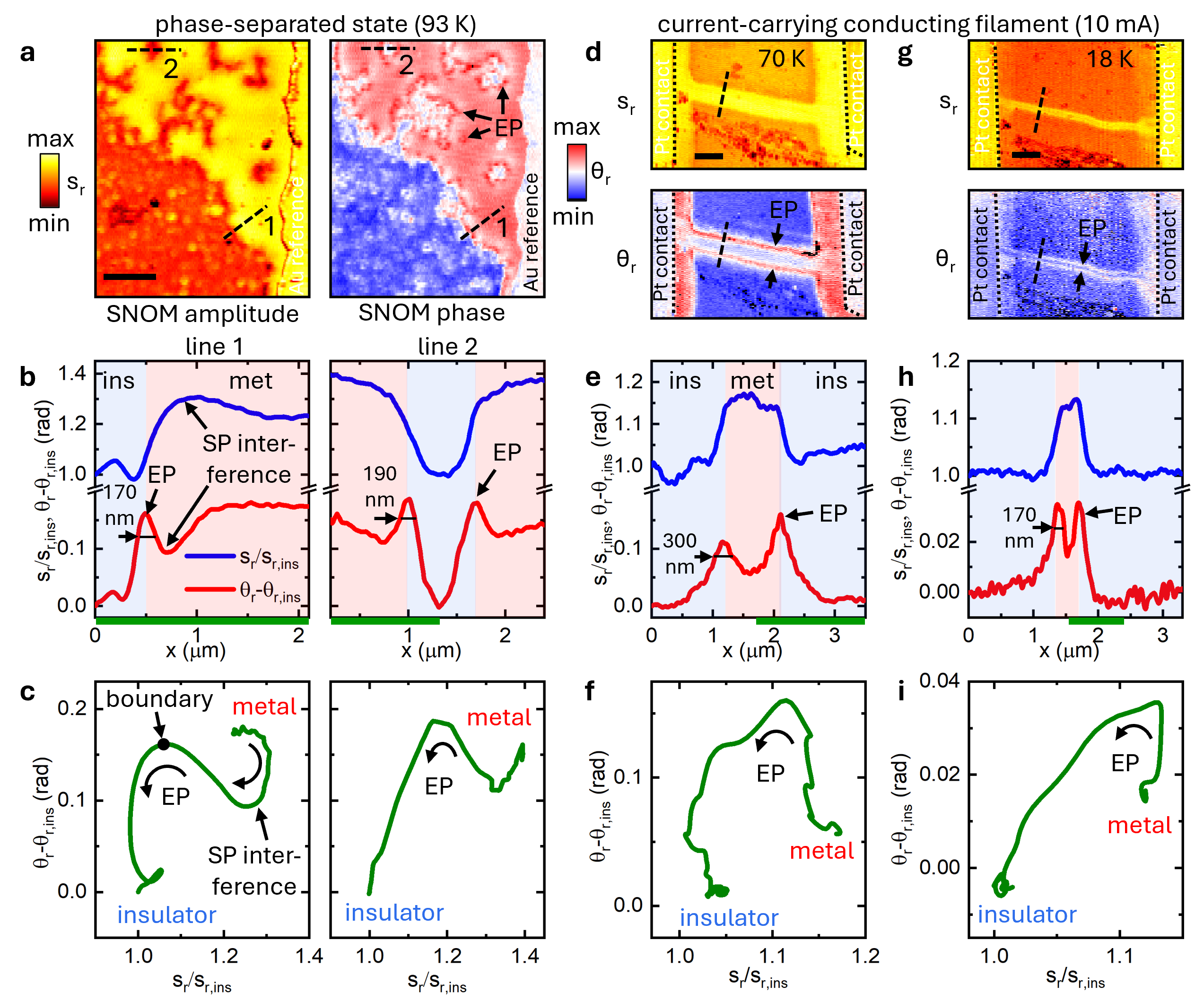}}
\caption{\textbf{Correlative amplitude-phase s-SNOM imaging of phase separated NdNiO$_{3}$ films.} \textbf{a}, Maps the amplitude $s_{r}$ (left) and phase $\theta_{r}$(right) of a 10 nm thick NNO film at 93~K. \textbf{b}, Line profiles along the dashed lines in \textbf{a}, normalized to their values of the insulator state.  \textbf{c}, Amplitude-phase correlations (APCs) corresponding to the profiles in \textbf{b} taken in the regions marked by horizontal green lines in the abscissa axes. \textbf{d} and \textbf{g}, s-SNOM images of the near field amplitude (left) and phase  (right) of a 40 nm thick NNO film  at a base temperature of 70 K and 18 K, while a current of 10 mA is sent through the film. The current passes through a bright  conducting filament, where the local temperature is higher than the transition temperature. \textbf{e} and \textbf{h}, Normalized amplitude and phase profiles along the dashed line in \textbf{d} and \textbf{g}. \textbf{f} and \textbf{i}, APC profile taken from the region shown by the green line in the abscissa axis in panels \textbf{e} and \textbf{h}. Scale bars: 2 $\mu$m. }
\label{Fig_ExpProfiles}
\end{figure*}

\noindent Examination of the s-SNOM phase images (Fig. \ref{Fig_ExpProfiles}) reveals new and complementary information not present in the amplitude images (Fig. \ref{Fig1}d). Fig. \ref{Fig_ExpProfiles}a shows maps of the normalized amplitude $s_{r}$ (left panel) and the normalized phase $\theta_{r}$ (right panel) of the same 10 nm thick NNO sample in the inhomogeneous state at 93 K. While both quantities are sensitive to the local film conductivity (being higher in the metallic regions), the s-SNOM phase shows a maximum directly at the boundaries (marked with solid arrows in the right panel), in contrast to the s-SNOM amplitude image, which shows a broadened step-like transition from the insulating to the metallic state. Fig. \ref{Fig_ExpProfiles}b illustrates this effect along two selected paths indicated by dashed lines in Fig. \ref{Fig_ExpProfiles}a. Path 1 crosses a single MI boundary, and path 2 goes through a small (about 600 nm in diameter) insulating island inside the metallic environment. At every boundary, the amplitude profiles (blue) show a broadened step function shape, while the phase profiles (red) are marked with a strong maximum. As shown in the same figure, the full width at half maximum (FWHM) of the phase peak is about 170-190 nm, which is well above the s-SNOM resolution. s-SNOM amplitude and phase images collected at other temperatures (Supplementary Fig. S3) show the same effect. Importantly, the MI boundaries are not associated with any AFM topography structure (Supplementary Fig. S5), which excludes a crosstalk between the AFM and s-SNOM signals as a mundane explanation of the phase peak.


Apart from the edge anomaly, the profiles along line 1 in Fig. \ref{Fig_ExpProfiles}b reveal some variation of the amplitude and phase inside the metallic region. The maxima and minima in $s_{r}$ and $\theta_{r}$ are shifted with respect to each other, suggesting that they are not due to the sample inhomogeneity. Very similar structures have been observed close to the edge in a topological insulator material Bi$_{2}$Se$_{3}$ and attributed to the interference of overdamped surface plasmon polaritons (SPs) excited by the tip, and reflected from the boundary \cite{ChenNC22}. As will be shown below, this is also the case in our samples.

Plasmonic features in the complex-valued s-SNOM data are identified most conveniently via analysis of the amplitude-phase correlation (APC) \cite{SchnellNC14,ChenNC22}. Fig. \ref{Fig_ExpProfiles}c presents a parametric plot of $\theta_{r}-\theta_{r,\text{ins}}$ as a function of $s_{r}/s_{r,\text{ins}}$ along the regions marked by horizontal green lines in the abscissa axes in Fig. \ref{Fig_ExpProfiles}b for path 1 (Fig. \ref{Fig_ExpProfiles}c left) and path 2 (Fig. \ref{Fig_ExpProfiles}c right). For both paths, the APCs have a peculiar 'flipped-S' shape made of two parts. The first one, where the curve rotates clockwise, stems from the mentioned non-monotonic variation of the amplitude and phase inside the metallic region. The second part, where the APC turns counterclockwise, corresponds to the MI crossover regions in Fig. \ref{Fig_ExpProfiles}b featuring the phase peak. As will be proven later, the first part of the APC can be attributed to the interference of SPs, emitted by the s-SNOM tip inside metallic regions and reflected from the MI boundaries, while the second part corresponds to the excitation of edge polaritons (EPs), when the tip is exactly above the boundary.

To demonstrate that edge phase anomaly is not specific to the case where the MI boundaries are spontaneously formed, we next consider a boundary generated via a different mechanism, namely, by passing a high electric current through a film at a base temperature $T_{b} < T_{\text{MIT}}$. In this process, a conducting filament comprising most of the current is deterministically formed, where the local temperature exceeds $T_{\text{MIT}}$ \cite{LuibrandPRR23}. The properties of the filaments depend strongly on the base temperature: for the same amount of the total current, the filaments are narrower for lower values of $T_{b}$ (See Fig. \ref{Fig_ExpProfiles}d and g). In Fig. \ref{Fig_ExpProfiles}d and g, we show maps of $s_{r}$ and $\theta_{r}$ measured at 70 K and 18 K respectively. In both cases, the same current of 10 mA is sent through a 40 nm thick NNO film deposited on a (001)-oriented LAO substrate (see Methods), between two Pt electrodes (marked with the dotted lines) evaporated on top of the film and separated by 10 $\mu$m from each other. Although the filament is much wider at 70 K, the same qualitative picture is observed at both temperatures: the amplitude image shows a uniform near-field intensity inside the filament, while the phase images are marked by sharp auras at the edges (indicated with arrows). The corresponding spatial profiles of $s_{r}$ and $\theta_{r}$ across the filament are given in Fig. \ref{Fig_ExpProfiles}e and h for 70 K and 18 K respectively. While the amplitude shows a nearly rectangular shape, the phase is marked with two outstanding peaks located exactly at the MI boundaries, resembling strikingly the case of phase separated state close to $T_{\text{MIT}}$ without electric current (Fig. \ref{Fig_ExpProfiles}b).

An important observation is that the phase peak is about two times narrower at low temperature (FWHM is 170 nm at 18 K vs. 300 nm at 70 K). This suggests that the width of the phase peak in this case is related to the physical width of the MI boundary, which is driven by the temperature variation across the edge \cite{LuibrandPRR23}. In our experiments, the narrowest phase peak (FWHM = 115 nm) has been detected in a different device with the same film thickness at 13 K (Supplementary Fig. S4). The APC curves for the three considered cases are shown in Fig. \ref{Fig_ExpProfiles}f, i and Supplementary Fig. S4c. They all reveal a counterclockwise curvature, typical for the EP anomaly. Overall, these data unequivocally demonstrate the excitation of edge-polaritons at the MI boundaries regardless the method used to trigger the MIT.

\hspace{2pt}

\noindent \textbf{\large Polariton dispersion and s-SNOM signal far from MI boundaries}

\noindent For a better understanding of the experimental polaritonic effects in the NNO thin films, we calculate the frequency-momentum dispersion of surface polaritons in an infinite uniform sample by means of the Fresnel reflection coefficient as the film metallicity evolves. For simplicity, we assume that the dielectric function $\varepsilon(\omega)$ can be linearly interpolated between the two regimes using a single metallicity parameter $f$: 
\begin{equation}\label{Eq1}
\varepsilon(f,\omega)=(1-f)\cdot\varepsilon_{\text{ins}}(\omega)+f\cdot\varepsilon_{\text{met}}(\omega).
\end{equation}

\noindent Physically, $f$ can be related to the Ginzburg-Landau order parameter describing the system symmetry change across the transition. In the nickelates, where the MIT involves charge disproportionation and structural distortion, it is common to associate the main order parameter with the difference between the size of the NiO$_{6}$ octahedra with low and high charge on the nickel atoms \cite{CatalanoRPP18,GawrylukPRB19}. Although the NNO also undergoes a magnetic ordering at the same temperature that is often described with an additional order parameter \cite{LeePRL11,PostNP18,RuppenPRB17}, for the purpose of this paper we use a single-parameter interpolation to avoid unnecessary complexity.

In Fig. \ref{Fig_Sim_Bulk}a-c, we present the calculated (as described in the Supplementary Note 1) color maps of the imaginary part of the momentum ($q$)- and frequency ($\omega$)- dependent Fresnel reflection coefficient $r_{p}(q,\omega)$ for $f=0$ (insulator), $f=f_{0}=0.16$ (where $\text{Re}(\varepsilon(f_{0},\omega_{\text{las}})=0$, corresponding to the ENZ regime) and $f=1$ (metal) respectively. The maxima of $\text{Im} (r_{p}(q,\omega))$ (bright regions) correspond to the SP dispersion. In the insulating state (Fig. \ref{Fig_Sim_Bulk}a), strong surface phonon-polaritons (SPhP) modes in the LAO substrate are observed below 750 cm$^{-1}$ (the dotted line follows the mode frequencies as a function of $q$). As the NNO metallicity increases, plasmon polaritons emerge in the films (Figs. \ref{Fig_Sim_Bulk}b and c), and SPhPs from the substrate couple with them forming hybrid surface plasmon phonon-polaritons that we shall shortly call surface polaritons. As a result of this coupling, the mode group velocity, $v_{g}=\partial\omega/\partial q$, increases from zero at $f=0$ to about 10\% of the speed of light at $f=1$ at $\omega_{\text{las}}$. In the ENZ regime (Fig. \ref{Fig_Sim_Bulk}b), the dispersion curve asymptotically approaches $\omega_{\text{las}}$ in the limit $q\rightarrow\infty$.


In the monochromatic s-SNOM experiment, the tip provides a broad distribution of momenta $q$ at a single frequency $\omega_{\text{las}}$ with the strongest tip-sample coupling at $q=q_{\text{tip}}=1/a$, where $a$ is the tip radius \cite{FeiNL11,MesterNC20}. Hence, the s-SNOM signal is expected to be mostly determined by the Fresnel coefficient $r_{p,\text{opt}}=r_{p}(q_{\text{tip}}, \omega_{\text{las}})$. Brown circles in Fig. \ref{Fig_Sim_Bulk}a-c represent this `optimal' set of $q$ and $\omega$ for the experimental value $a$ = 40 nm. One can see that the SP branch evolves in such a way that it almost matches the optimal point at $f=f_{0}$. To further develop on this, we present in Fig. \ref{Fig_Sim_Bulk}d the real (blue) and imaginary (red) parts of $r_{p,\text{opt}}$ as a function of $f$. As the metallicity increases, $\text{Re}(r_{p,\text{opt}})$ initially shows a small decrease, then suddenly raises with the steepest slope at $f\approx f_{0}$ and then remains almost constant up to $f=1$. In contrast, $\text{Im}(r_{p,\text{opt}})$ is small both at $f=0$ and $f=1$ but it features a strong maximum in the ENZ regime due to the mentioned momentum-frequency matching effect. Interestingly, the parametric plot of the imaginary part vs the real part of $r_{p,\text{opt}}$ (inset) shows an almost perfectly circular arc with the ENZ regime being exactly in the middle.  

Based on the calculated SP dispersion $r_{p}(q,\omega)$, we next compute the s-SNOM signal far from the boundaries using an oscillating point-dipole-approximation (Fig. \ref{Fig_Sim_Bulk}e). In this model, the s-SNOM tip is replaced by a polarizable point dipole with the vertical ($z$-axis) polarizability equal to that of a sphere with the radius $a$ and with a negligible horizontal polarizability. The dipole mechanically oscillates along the $z$-axis mimicking the tapping regime, with the peak-to-peak amplitude $z_{\text{amp}}$ and the lowest dipole position $z_{\text{min}}$ matching the experimental values (in our case 80 and 40 nm respectively). Both the tip and sample are illuminated by a monochromatic plane wave from the laser source incident at an angle close to the experimental one (45$^{\circ}$). The complex s-SNOM signal is proportional to the induced dipole moment, which depends on the dipole height, and the tapping harmonics are obtained via a Fourier transform of its $z$-dependence (see Supplementary Note S3). The resulting amplitude and phase of the s-SNOM signal (normalized to the insulator-state values) as a function of $f$ are shown in Fig. \ref{Fig_Sim_Bulk}f. The two curves resemble strikingly the $f$-dependence of the real and imaginary parts of $r_{p,\text{opt}}$ (Fig. \ref{Fig_Sim_Bulk}d), with a notable phase-peak in the ENZ regime. The corresponding APC (inset) also forms a circular arc. The obvious correlation between Fig. \ref{Fig_Sim_Bulk}d and \ref{Fig_Sim_Bulk}f allows us to conclude that the phase-peak observed as a function of $f$ is caused by the enhanced near-field absorption in the ENZ regime due to resonant excitation of the SPs by the s-SNOM tip. We note that the model APC parametrized by $f$ (inset in Fig. \ref{Fig_Sim_Bulk}f) rotates counterclockwise while evolving from the metallic to the insulating state, similar to the experimental APC parametrized by the tip position (Fig. \ref{Fig_ExpProfiles}c, f and i). 

\hspace{2pt}

\noindent \textbf{\large Edge polaritons and the nature of the phase peak}

\noindent Although it is tempting to ascribe the s-SNOM phase peak observed at the MI boundaries to the spatial variation of the metallicity parameter $f$ across the boundary and therefore to the ENZ-related phase maximum (Fig. \ref{Fig_Sim_Bulk}f), this explanation alone faces a significant problem. Specifically, it requires the spatial variation of $f$ to occur on the scales exceeding the SP wavelength $\lambda_{\text{SP}}=2\pi/\text{Re}(q_{\text{SP}})$. However, in the fully metallic state, the SP branch crosses the s-SNOM frequency line at $q_{\text{SP}}\approx 0.15\times 10^5 $ cm$^{-1}$ (Fig. \ref{Fig_Sim_Bulk}c), resulting in $\lambda_{\text{SP}}\approx$ 4~$\mu$m, which exceeds dramatically the observed MI boundary width of below 100-150 nm (Fig. \ref{Fig_ExpProfiles}, see also Ref. \cite{PostNP18}). This suggests that the actual picture is more complex.


To better understand the behavior of the s-SNOM signal at a boundary and how it depends on the boundary smoothness, we resort to finite-element method (FEM) simulations (see Methods), assuming that the order parameter varies with $x$ in the following way:

\begin{equation}\label{Eq2}
    f(x) = \begin{cases}
        \Theta(x), &\text{if } |x| > w/2 \\
        x/w + 1/2, &\text{if } |x| \leq w/2
    \end{cases}
\end{equation}

\noindent where $\Theta(x)$ is the Heaviside step-function and $w$ is the physical width of the boundary (centered at $x=0$). To simplify the simulations, we adopt an approximation, where the dipole, which is a primary field emitter, is kept at a given height $z_{d}$ above the sample while moving along the surface. Furthermore, the s-SNOM signal is replaced by the $z$-axis component of the electric field, $E_{z}$, measured in a probe point below the dipole at the height $z_{p}$ above the surface (which we fix to 25 nm), as sketched in the inset of Fig. \ref{Fig_Sim_Edge}a. This approximation has been shown to work well to describe the behavior of the s-SNOM signal close to sample edges in various materials \cite{NikitinNPh16, AlvarezAM20, ChenNC22}. In Supplementary Note 3 we also demonstrate that in the case of the s-SNOM signal far from edges, this simplified approach matches well the results for the complete s-SNOM model.

The key question to be answered is how much the FEM-simulated field $E_{z}(x_{d})$ as a function of the horizontal dipole coordinate $x_{d}$ differs from the \textit{gedanken} `local' value $E_{z,\text{loc}}(x_{d})$, obtained by assuming that the film is uniform (no metal-insulator boundary) with the metallicity equal to $f(x_{d})$ and the dielectric function of $\varepsilon(f(x_{d}),\omega)$. In the limit $w \gg \lambda_{\text{SP}}$, the two must obviously coincide, but if the boundary width is comparable to or especially much lower than the wavelength then such match is no longer expected.

In Fig. \ref{Fig_Sim_Edge}a and b we present two FEM simulations, where $w$ is set to 1 $\mu$m (smooth boundary) and 0 (sharp boundary) respectively. In both cases, the dipole height is $z_{d}$=200 nm, which is much smaller than the plasmon wavelength. The blue (red) curves indicate the position dependent amplitude (phase) of the actual field normalized to the one in the case of a uniform insulator $E_{z}(x_d)/E_{z,\text{ins}}$. The corresponding amplitude and phase of the local value $E_{z,\text{loc}}(x_d)/E_{z,\text{ins}}$ are shown with dashed dotted lines. For both values of $w$, the amplitude shows a broadened step-like behavior while the phase features with a strong peak, in qualitative agreement with the experiments. A closer look, however, reveals substantial differences between the two cases. For the smooth boundary (Fig. \ref{Fig_Sim_Edge}a), the actual and the local values almost coincide. This means that the shape (and the width) of the phase peak is determined by the variation of $f$ and therefore the origin of the peak is the ENZ-enhanced absorption in the area of the boundary where $f(x_{d})\approx f_{0}$. In contrast, in the case of a sharp boundary (Fig. \ref{Fig_Sim_Edge}b), the real and the local values of both amplitude and phase are totally different. Specifically, the local value has the step-function shape, which rules out the ENZ origin of the phase peak, implying that another mechanism is in action. 

The difference in the behavior of the s-SNOM signal on smooth and sharp boundaries is further supported by the analysis of the amplitude-phase correlations (Fig. \ref{Fig_Sim_Edge}c). For $w = 1$ $\mu$m (brown line) the APC shows an incomplete circle almost matching the local APC (dashed-dotted line) and similar to the one shown in the inset of Fig. \ref{Fig_Sim_Bulk}f. In contrast, the APC for the sharp edge (pink line) shows an outstanding kink separating the metal and insulator segments. Interestingly, these two qualitatively distinct behaviors crossover one into another by a continuous change of the boundary width. The green curve in Fig. \ref{Fig_Sim_Edge}c refers to an intermediate value $w$=100 nm (the corresponding position dependence is given in the Supplementary Fig. S6). This APC clearly deviates from the local curve, however the kink is smoother as compared to the sharp edge. We note that the shape of the green curve strikingly resembles the experimental APCs across metallic filaments (Fig. \ref{Fig_ExpProfiles}f,i), suggesting that in reality the intermediate situation is likely to occur.


To establish the physical origin of the phase peak anomaly in the case of sharp boundary, we calculate the $xy$ map of the squared norm of the vertical field, $|E_{z}(x,y,z=z_{p})|^2$, when the dipole is located 200 nm above the center of the MI boundary. Fig. \ref{Fig_Sim_Edge}d and e present the cases of the smooth and the sharp boundaries respectively. In both cases, the field distribution deep inside the metal and the insulator regions is similar to the one when the dipole is on top of a pure metal and pure insulator respectively (Supplementary Fig. S7). However, the two maps are very different near the edge. Unlike the first case, a well defined edge state (ES) is present on the sharp boundary as shown by the black arrows. While the ES is very narrow, of the order of 100 nm (inset of Fig. \ref{Fig_Sim_Edge}f shows a  crosscut 2.5 $\mu$m away from the dipole), it extends several microns away from the source in the center, clearly demonstrating its one-dimensional character and implying that the electromagnetic energy radiated by the dipole is canalized by the ES (Extended Data contains two animations \href{https://drive.google.com/file/d/1p5-g72AdZCs-dv9BROliIEbNESZ9zKeY/view?usp=drive_link}{AnimationSmoothEdge.gif} and  \href{https://drive.google.com/file/d/1X1vpq7vZIU4h5k0dQ0ZAyL3zsiH76X31/view?usp=drive_link}{AnimationSharpEdge.gif} showing how the field distribution maps in the cases of the smooth and sharp boundaries evolve as the dipole moves across them). The phase maximum observed in s-SNOM can therefore be attributed to the extra dissipation of energy along the ES when the tip is sufficiently close to the edge. 

To see how the ENZ and the ES scenarios cross over to each other, in Fig. \ref{Fig_Sim_Edge}f we show field profiles for various widths along the line crossing the boundary 2.5 $\mu$m away from the center (black dashed lines in Fig. \ref{Fig_Sim_Edge}d and e), where the direct dipole field is negligible. For the sharp boundary, Fig. \ref{Fig_Sim_Edge}e, the ES manifests as a asymmetric sharp peak at $x=0$, decaying faster toward the insulating region. The peak is also present for the 100 nm wide boundary, though it is half as intense and shifted approximately about 40 nm ($=0.4 w$) left from the center. For the smooth boundary ($w=1\,\mu$m), there is also a maximum shifted by about 40 percent of the width to the left. Although this peak is about twenty times weaker than the one for the sharp edge, it presents a field still significantly higher than the field above a pure metal (no metal-insulator interface, blue curve), which means that even in the very smooth boundary there is some canalization of energy along the edge even in the ENZ limit. Interestingly in both cases of $w$ = 100 nm and 1 $\mu$m, the maximum is located at the $x$ position where $f=f_{0}$, signaling that most of the energy flows in ENZ region of the film. 

\hspace{2pt}

\noindent \textbf{\large SP interference at the MI boundaries}

\noindent As discussed earlier, the s-SNOM amplitude and phase profiles exhibit extra structure inside the metallic region, where the APC vector rotates clockwise (Fig. \ref{Fig_ExpProfiles}b,c). This phenomenon, which was widely observed in various polaritonic materials \cite{ChenNature12,FeiNature12,DaiNN15,MaNature18,ChenNC22}, stems from the constructive and destructive interference between the tip illuminating field and the one due to excited and edge-reflected SPs (inset in Fig. \ref{Fig_Sim_Edge_2}b). 

It is clear from Fig. \ref{Fig_Sim_Bulk}c that the SP momentum at the laser frequency $q_{\text{SP}}$ is about 20 times smaller than $q_{\text{tip}}$, making its excitation by a 40 nm sharp s-SNOM tip quite inefficient. This explains why there is either no signature or a very weak indication of the SP interference in the simulation for the low position of the dipole emitter (Fig. \ref{Fig_Sim_Edge}b,c). On the other hand, the actual tip is not a point dipole but rather a several-micron high pyramidal antenna sharpened at the apex. The tip-SP momentum mismatch can be therefore less prohibiting in reality than in the point dipole model that we use. To verify this, in Fig. \ref{Fig_Sim_Edge_2}a-c we present another simulation, where all parameters are the same as in the previous one (Fig. \ref{Fig_Sim_Edge}a-c) except $z_{d}$, which we now set to 1.5 $\mu$m, much closer to the SP wavelength (about 4 $\mu$m) than the value of 200 nm used before. Now the signature of the SP interference becomes obvious both for the smooth boundary, $w$= 1 $\mu$m, and for the sharp boundary, $w=0$. In the latter case the structure is expectedly more pronounced, since the sharp boundary reflects the SP better than the smooth one. Interestingly, the sharp edge plasmon structure is seen very clearly also for the high-dipole position, and it has, moreover about the same width as for $z_{d}$ = 200 nm. The APC curves for the smooth, intermediate ($w=100\,\text{nm}$) and sharp boundaries are given in Fig. \ref{Fig_Sim_Edge_2}c, where the SP interference manifests as a clockwise rotation of the state vector in the metallic state. Specifically, the intermediate case (green curve) strongly resembles the experimental APC in the phase-separated state (Fig. \ref{Fig_ExpProfiles}b,c). Given that substantial approximations has been adopted in the used model, we do not pursue a close quantitative matching between experiment and theory by fine tuning the tip and the sample parameters. Nevertheless, the striking qualitative agreement makes the presence of both edge- and surface polaritons in the s-SNOM data sufficiently obvious.

\hspace{2pt}

\noindent \textbf{\large Discussion}

\noindent In this work, we have observed well-defined polaritonic features in a material where plasmonic effects are generally expected to be weak \cite{RutaScience25}, in particular, at the edges of the regions between metallic and insulating states formed during a metal-insulator transition. In the limit of a sharp boundary, these quasi-one dimensional modes canalize the flow of electromagnetic energy to a narrow edge state with the cross-section of about 100 nm (see the inset in Fig. \ref{Fig_Sim_Edge}f) over distances of several microns (Fig. \ref{Fig_Sim_Edge}e). The ratio between the propagation length and the lateral field spread, which is one of the most important plasmonic figures of merit \cite{CaldwellNanophotonics15}, is well above 10 in this case, demonstrating the potential usefulness of this phenomenon in plasmonics. Such a long propagation is possibly related to the fact that most of the field is present in vacuum, which reduces the losses as compared to the usual SPs \cite{NikitinPRB11,ZhaoOE16}. In this regard, it is essential that the film used in this study is only several nanometers thick, bringing the edge state close to one-dimensional limit. Moreover, the small thickness is important for the plasmon-phonon hybridization, which results in a stronger confinement compared to surface plasmon-polaritons in a bulk sample.

We have theoretically shown that the origin of the phase peak is two-fold, being driven by the enhanced excitation of the SPs when the s-SNOM tip is above an ENZ region of the sample inside a smooth boundary or by launching of edge polaritons when the tip is above the edge state at a sharp edge. Interestingly, these two qualitatively different regimes (ENZ and ES) continuously connect between each other as the physical boundary width changes. It is therefore legitimate to speak about a generalized edge polariton (defined as a polariton observed only at the edge), the electromagnetic structure of which evolves with the edge smoothness.

The width of the phase peak can serve as indicator of the regime (ENZ or ES or a combination of the two) being actually realized. Our simulations show that the peak is broader in the smooth boundary (FWHM = 400 nm in Fig. \ref{Fig_Sim_Edge}a), where it is proportional to the boundary width. On the other hand, the peak width for the sharp edge (170 nm in Fig. \ref{Fig_Sim_Edge}a) is limited by the field distribution around at the boundary, which is indeed of the order of 100 nm (Fig. \ref{Fig_Sim_Edge}f). By comparing with the experiment (Fig. \ref{Fig_ExpProfiles}), we conclude that in the phase separated phase close to $T_{\text{MIT}}$, the peak width ($<$ 200 nm) is close to the ES limit. On the other hand, for the current-carrying filament at 70 K, the broad peak width (FWHM = 300 nm) indicates that the MI interface is more smeared as compared to the data at 18 K (170 nm) and 13 K (115 nm).

To conclude, the plasmonic effects influence strongly the s-SNOM maps at metal-insulator boundaries even in "bad metals" by making them in general "non-local". Therefore,  ignoring them may result in a significant overestimation of the actual boundary width. On the other hand, analyzing the amplitude-phase correlations is a useful test for the applicability of the local interpretation of the near-field spatial profiles. The phase boundaries in materials undergoing MITs can be engineered using various approaches, such as thermal variation, substrate induced strain or electric current. Hence, the observation of robust plasmonic effects in the correlated functional oxides opens new avenues for tunable nanophotonics. 

\hspace{2pt}

\noindent \textbf{\large Methods}

\noindent \textbf{Sample fabrication.} Epitaxial films of NdNiO$_{3}$ (NNO) were grown by off-axis radio-frequency magnetron sputtering on (001) oriented LaAlO$_{3}$ (LAO) substrates and the desired thickness was verified using X-ray diffraction. A 10 nm thick film was used for temperature dependent SNOM measurements, where 40 nm thick Au reference areas were deposited by optical lithography. For the electric-current dependent measurements, a 40 nm thick NNO film was used for fabrication of microdevices, which involved both optical and e-beam lithography as described in Ref.\cite{LuibrandPRR23}.

\noindent \textbf{Far-field reflectivity measurements.} Temperature dependent infrared near-normal reflectivity spectra of the NNO/LAO samples were obtained using a Fourier transform infrared (FT-IR) spectrometer (Bruker Vertex 70v) equipped with an FT-IR microscope (Bruker Hyperion 2000) in a liquid-helium flow cryostat (CryoVac Konti-Micro). A 40 nm thick layer of gold was used as a reference to obtain the absolute reflectivity value. The dielectric function of the film $\varepsilon(\omega)$ was extracted by a Drude-Lorentz model fitting of the reflectivity data using a dielectric function of LAO known from a separate reflectivity measured on a pure substrate \cite{LuoNC19}.

\noindent \textbf{Near-field cryogenic s-SNOM measurements.} s-SNOM maps of amplitude, phase and topography at various temperatures were obtained using a commercial cryo-NeaSNOM system from Neaspec/Attocube. The method principle is described in the main text. A CO$_{2}$ laser and quantum cascades laser (QCL) were used as sources and a MCT detector was employed. 

\noindent \textbf{Electromagnetic simulations.} 3D electromagnetic finite-element method (FEM) simulations were performed by using COMSOL Multiphysics in a computational volume of 40$\times$40$\times$40 $\mu$m$^{3}$ surrounded by perfectly matching layers (PMLs) to avoid spurious reflections. A point-dipole monochromatic source was used to mimic the AFM tip. An optimal 3D grid was chosen to ensure a fast convergency and sufficient accuracy, and the results were checked to be independent on the further mesh thinning.
As the film thickness is $10^3$ smaller than the infrared wavelength (10 nm as compared to 10 $\mu$m), it was substituted with an equivalent 2D conducting layer having the same optical 2D optical conductivity as the film.

\hspace{2pt}

\noindent \textbf{\large Author contributions} 

\noindent ABK conceived the project idea and analyzed the data. WL and AB performed the measurements and analyzed the data. CD and JdV fabricated and characterized the samples. JT and JT-G contributed to the data analysis. ABK, WL and AB wrote the manuscript using input from all authors. Everybody contributed to discussions.

\hspace{2pt}

\noindent \textbf{\large Acknowledgments} 

\noindent WL was supported by the National Key Research and Development Program of China (2024YFA1409500) and the National Natural Science Foundation of China (12127803,12004196). AB, JT, JT-G and ABK acknowledge the support from the Swiss National Science Foundation (SNSF) via Research projects 200020\_201096 and 200021\_236697. JT-G was also supported by the SNSF Swiss Posdoctoral Fellowship (TMPFP2\_224378). JdV was financed by the Spanish Ministry of Science, Innovation and Universities via the ‘Ramón y Cajal’ grant RYC-2021-030952-I.

\bibliography{ref}
\newpage

\newpage


\newpage
\begin{figure*}[htb]
\centerline{\includegraphics[width=18cm]{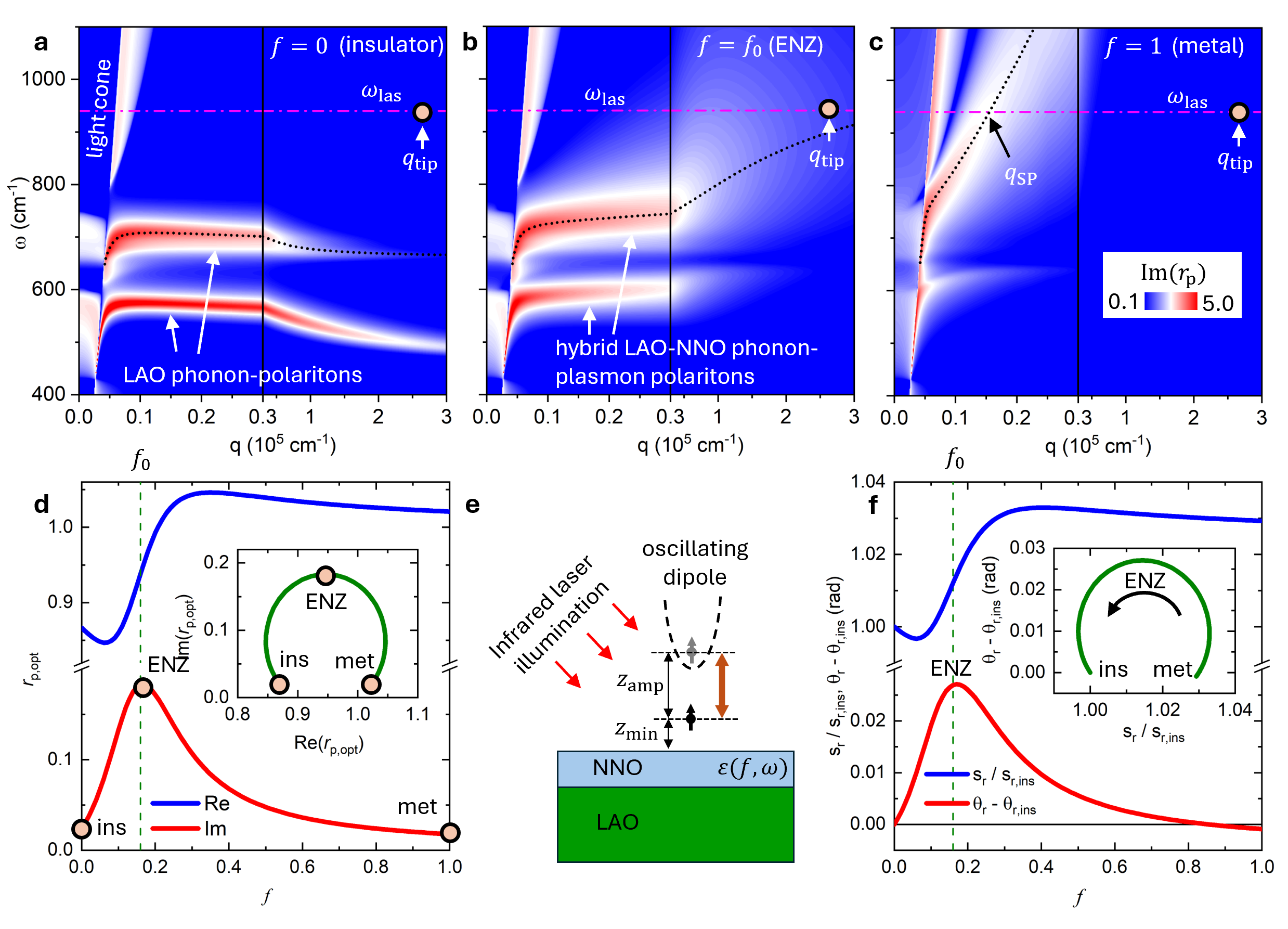}}
\caption{\textbf{Calculated surface polariton dispersion and s-SNOM signal far from boundaries as a function of the film metallicity.} \textbf{a}-\textbf{c}, Frequency-momentum maps $\text{Im}(r_{p}(q,\omega))$ for $f$ = 0 (insulator), $f_0$ (ENZ) and 1 (metal) respectively. The brown circles indicate the values of $q_{\text{tip}}=1/a=2.5\times 10^5 \text{ cm}^{-1}$ and $\omega_{\text{las}}$=940 cm$^{-1}$. Note the horizontal scales change at $q=0.3\times10^{5}$ cm$^{-1}$. \textbf{d}, Real and imaginary parts of $r_{p,\text{opt}}=r_{p}(q_{\text{tip}},\omega_{\text{las}})$ as a function of $f$ (the inset shows the parametric Im vs. Re. dependence). \textbf{e} Sketch of the model used to calculate the s-SNOM signals. \textbf{f}, insulator-normalized amplitude $s_{r}/s_{r,\text{ins}}$ (blue line) and phase  $\theta_{r}-\theta_{r,\text{ins}}$ (red line) of the s-SNOM signal, with the corresponding APC shown in the inset.}
\label{Fig_Sim_Bulk}
\end{figure*}

\newpage

\begin{figure*}[htb]
\centerline{\includegraphics[width=18cm]{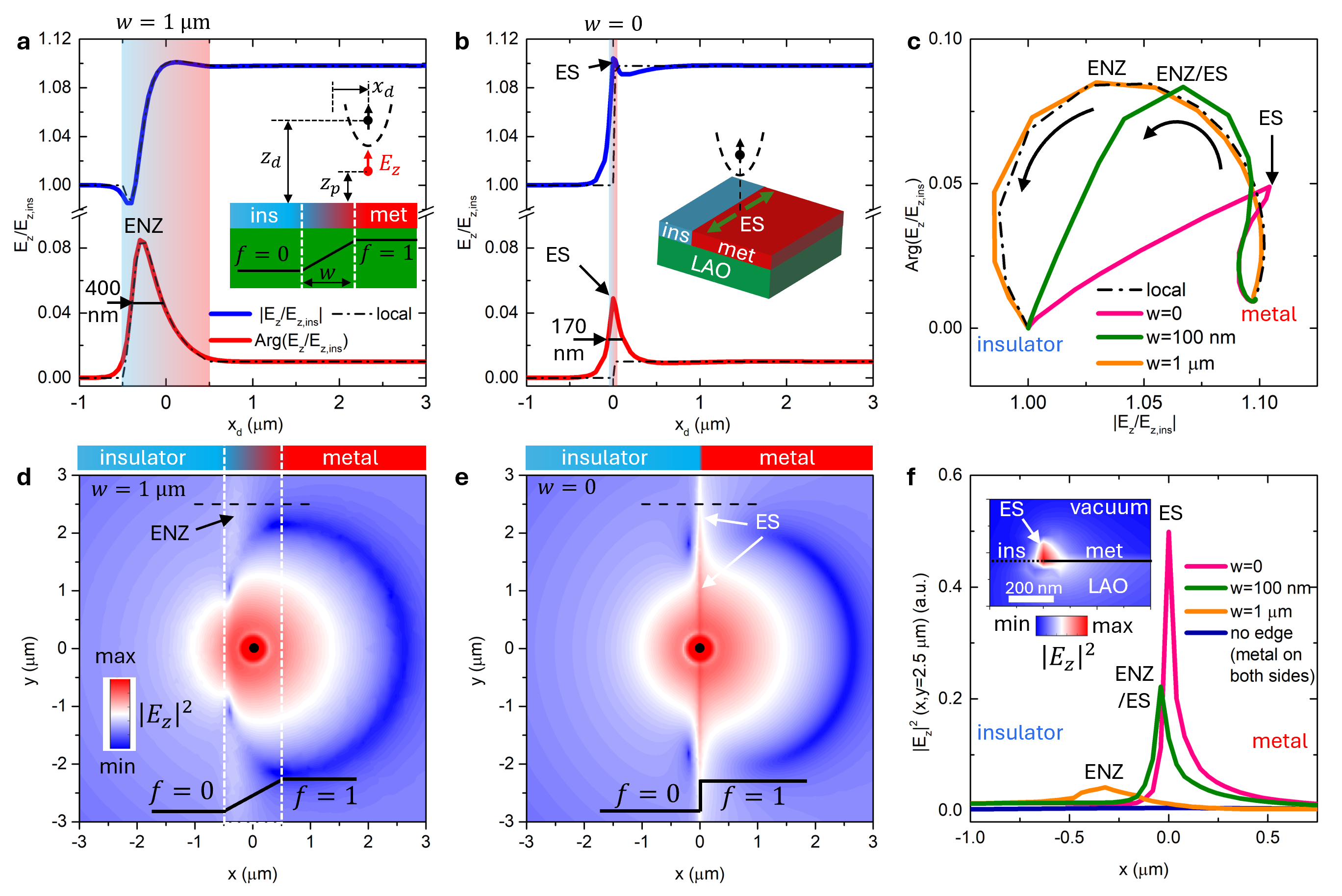}}
\caption{\textbf{Simulated s-SNOM signal across metal-insulator boundaries of various widths.} \textbf{a} and \textbf{b}, Amplitude (blue) and phase (red) of the probe-point field $E_{z}$ calculated on a smooth boundary (\textbf{a}) and a sharp boundary (\textbf{b}) as a function of the dipole position $x_{d}$ with respect to the center of the edge. The dipole height $z_{d}$ is 200 nm. For comparison, the dashed-dotted lines represent the corresponding values calculated far from edges on a sample with a uniform metallicity equal to the local value $f(x_{d})$. \textbf{c}, APC curves for the sharp ($w$=0), intermediate ($w$=100 nm) and smooth ($w$ = 1 $\mu$m) boundaries compared with the local curve (dashed-dotted line). \textbf{d} and \textbf{e}, The $xy$ map (logarithmic color scale, the same for both panels) of the electric field, $|E_{z}(x,y,z)|^2$, at the level $z$ = 25 nm above the surface, when the dipole emitter is located above the smooth (\textbf{d}) and the sharp (\textbf{e}) boundary ($x_{d}$=0, $z_{d}$ = 200 nm). \textbf{f}, Electric field profiles along the dashed line in panels (\textbf{d}) and (\textbf{e}) for the sharp, intermediate and smooth boundaries as well as for a uniform metallic sample without boundaries. The inset shows the distribution of $|E_{z}(x,y,z)|^2$ in the $xz$ plane (linear color scale) for the sharp edge at the same distance from the dipole emitter (2.5 $\mu$m).  The insets in \textbf{a} and \textbf{b} describe the calculation geometry and depict the excitation of the edge polaritons by the tip. In all simulations, $z_{p}$ = 25 nm.}
\label{Fig_Sim_Edge}
\end{figure*}

\newpage

\begin{figure*}[htb]
\centerline{\includegraphics[width=18cm]{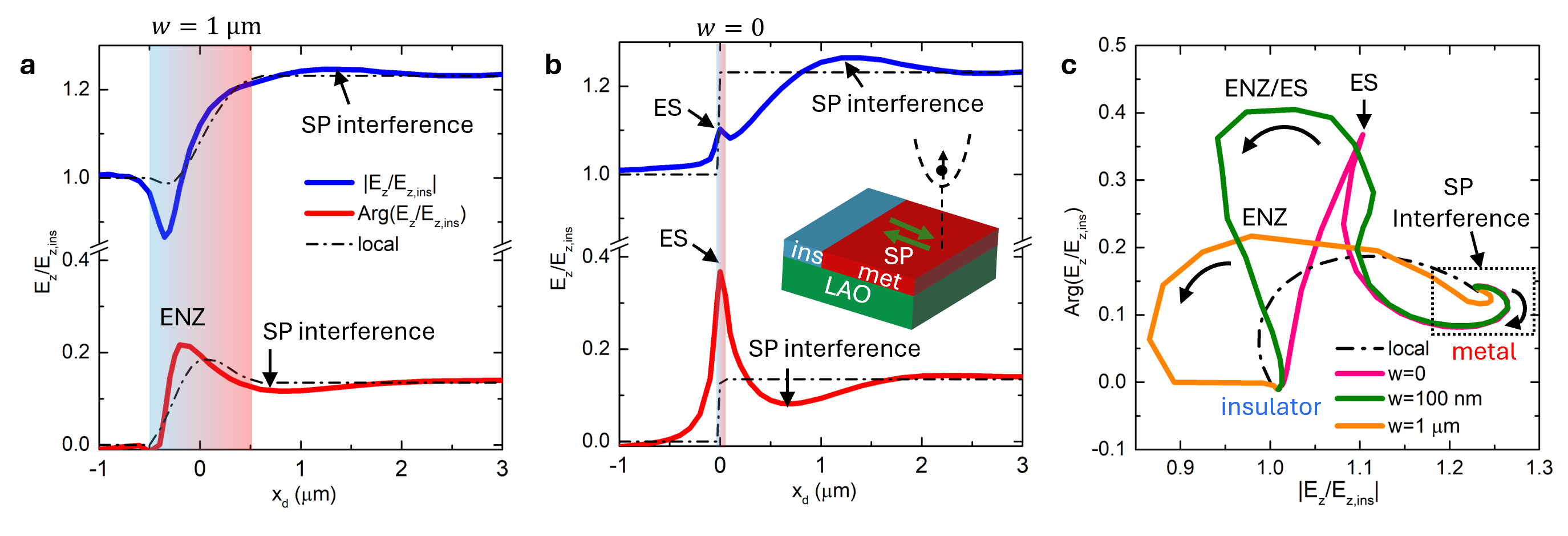}}
\caption{\textbf{Simulated s-SNOM signal across metal-insulator boundaries for a high dipole position.} \textbf{a}-\textbf{c}, The same as \textbf{a}-\textbf{c} in Fig. \ref{Fig_Sim_Edge}, but for a high dipole position $z_{d}$ = 1.5 $\mu$m. The inset in \textbf{b} depicts the excitation of SPs by the tip and their reflection from the boundary. In all simulations, $z_{p}$ = 25 nm.}
\label{Fig_Sim_Edge_2}
\end{figure*}

\end{document}


\date{\today}
\title{Supplementary Information: Edge polaritons at metal-insulator boundaries in a phase separated correlated oxide}

\author{Weiwei Luo}
\thanks{These authors contributed equally to this work.}
\affiliation{The Key Laboratory of Weak-Light Nonlinear Photonics, Ministry of Education, School of Physics and TEDA Applied Physics Institute, Nankai University, Tianjin 300457, China}
\affiliation{Department of Quantum Matter Physics, University of Geneva, Geneva 1211, Switzerland}

\author{Adrien Bercher}
\thanks{These authors contributed equally to this work.}
\affiliation{Department of Quantum Matter Physics, University of Geneva, Geneva 1211, Switzerland}

\author{Claribel Dominguez}
\affiliation{Department of Quantum Matter Physics, University of Geneva, Geneva 1211, Switzerland}

\author{Javier del Valle}
\affiliation{Department of Physics, University of Oviedo, C/ Federico García Lorca 18, 33007 Oviedo, Spain}
\affiliation{Center of Research on Nanomaterials and Nanotechnology, CINN (CSIC-Universidad de Oviedo), El Entrego 33940, Spain}

\author{J\'{e}r\'{e}mie Teyssier}
\affiliation{Department of Quantum Matter Physics, University of Geneva, Geneva 1211, Switzerland}

\author{Javier Taboada-Gutiérrez}
\affiliation{Department of Quantum Matter Physics, University of Geneva, Geneva 1211, Switzerland}

\author{Alexey B. Kuzmenko}
\email{Alexey.Kuzmenko@unige.ch}
\affiliation{Department of Quantum Matter Physics, University of Geneva, Geneva 1211, Switzerland}

\maketitle

\hspace{2pt}

\noindent \textbf{\large Note 1. Calculation of the Fresnel coefficient} 

\hspace{2pt}

\noindent For a film of thickness $d$ on an infinitely thick substrate, the Fresnel reflection coefficient (for the p-polarized modes) is equal to: 
\begin{equation}
  r_p(q,\omega)=\frac{r_{1,2}+r_{2,3}e^{2i k_{2}d}}{1+r_{1,2}r_{2,3}e^{2i k_{2}d}},\label{EqRp}
\end{equation}

\noindent where $r_{i,j}=(k_{i}\varepsilon_{j}-k_{j}\varepsilon_{i})/(k_{i}\varepsilon_{j}+k_{j}\varepsilon_{i})$, $k_i=\sqrt{\varepsilon_{i}q_0^2-q^2}$, $q_{0}=\omega/c$ and $q$ is the in-plane momentum. In our case, the indices 1, 2 and 3 in Eq. (\ref{EqRp}) refer to vacuum ($\varepsilon = 1$), the film ($\varepsilon = \varepsilon_{\text{NNO}}$) and the substrate ($\varepsilon = \varepsilon_{\text{LAO}}$) respectively.

\hspace{2pt}

\noindent \textbf{\large Note 2. Extraction of the dielectric function spectra of NdNiO$_{3}$} 

\hspace{2pt}

\noindent In order to obtain the spectra of the dielectric function of NdNiO$_{3}$ films, we fitted the broad-band normal-incidence reflectivity spectra (Fig. 1b) using a Drude-Lorentz model, with the parameters changing as a function of temperature. In the modeling, the reflectivity was calculated as $R= |r_p(q=0,\omega)|^2$, with $r_p(q,\omega)$ given by Eq. (\ref{EqRp}). The dielectric function of LaAlO$_3$, needed for this analysis has been extracted in a prior work \cite{LuoNC19}. The resulting spectra of  $\varepsilon_{\text{NNO}}(\omega)$ are shown in Fig. S1. The Drude behavior is obviously dominating the high-temperature spectra.

\begin{figure*}[htb]
\centerline{\includegraphics[width=12cm]
{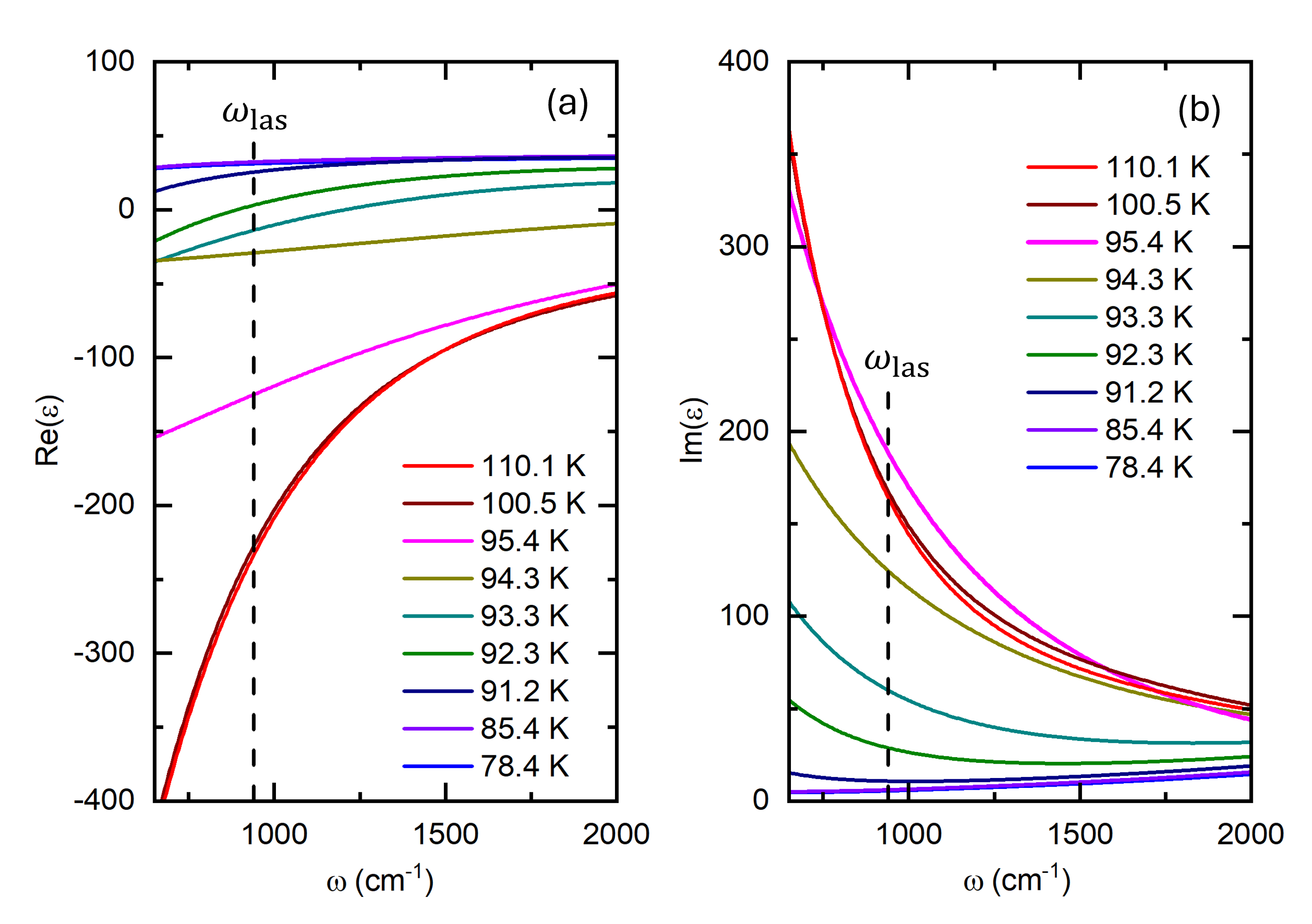}}

\begin{justify}
\textbf{Fig. S1.} Spectra of the real (a) and imaginary (b) parts of the dielectric function of NdNiO$_{3}$ at different temperatures at cooling down, extracted from the reflectivity spectra (Fig. 1b). The dashed line denotes the s-SNOM laser frequency.
\end{justify}

\label{Fig_SuppEpsilon}
\end{figure*}

\noindent

\hspace{2pt}

\noindent \textbf{\large Note 3. Analytical model for the s-SNOM signal far from boundaries} 

\hspace{2pt}

\noindent To calculate the s-SNOM signal, we replace the AFM tip with a point dipole located at a distance $z_{d}$ from the sample surface. The sample is supposed to be translationally invariant and optically isotropic in the $xy$ plane, in which case its electromagnetic properties are entirely described by the complex momentum- and frequency- dependent Fresnel reflection coefficient $r_{p}(q,\omega)$ (see Note 1). The dipole is assumed to be polarizable only in one (vertical) direction so that its dipole moment is always oriented along the $z$-axis and is proportional to the vertical component of the electric field at the dipole location:
\begin{equation}
d=\alpha E_{d,z}\label{Eqdz}.
\end{equation}
\noindent It is convenient to represent the polarizability via the effective radius $a$: $\alpha=4\pi \varepsilon_{0} a^3$, where $\varepsilon_{0}$ is the vacuum permittivity, (we are using the SI units). The physical meaning of this value is that the polarizability of a sphere of radius $a$ made of a material with a dielectric function $\varepsilon_{s}$ is $4\pi \varepsilon_{0}(\varepsilon_{s}-1)/(\varepsilon_{s}+2)a^{3}$. For a noble metal used for coating the SNOM tip, $\varepsilon_{s}$ has a large negative real part and a small positive imaginary part, so that $(\varepsilon_{s}-1)/(\varepsilon_{s}+2)\approx 1$. Therefore, $a$ should be of the order of the actual tip radius (about 40 nm).

Within a complete s-SNOM model (Fig. 3e from the main text and Fig. S2a), the dipole and the sample are illuminated from a remote source by a monochromatic p-polarized plane wave of the amplitude $E_{0}$, frequency $\omega$ and the angle of incidence $\theta$ . All electromagnetic quantities (fields, induced polarizations, dipole moments) oscillate in time proportionally to $e^{-i\omega t}$ and in the following, the time dependent factor will be omitted in the formulas. 

The field at the dipole location consists of the three contributions:
\begin{equation}
    E_{d,z}=E_{\text{inc},z}+E_{\text{ref},z}+E_{\text{ima},z}\label{EqEdz}
\end{equation}
\noindent Here 
\begin{equation}
   E_{\text{inc},z}=E_{0} \sin{\theta} e^{-i q_{0}z_{d} \cos{\theta}}\label{EqEincz}
\end{equation}
\noindent is the field of the incident light (direct illumination) and
 \begin{equation}
       E_{\text{ref},z}=r_{p}(q_{\text{inc}},\omega) E_{0}\sin{\theta} e^{i q_{0}z_{d} \cos{\theta}}\label{EqErefz}
\end{equation}
\noindent is the one of the reflected light (indirect illumination), where $q_{0}=\omega/c$ is the light wavevector in free-space and $q_{\text{inc}}=q_{0}\sin{\theta}$ is its projection on the $xy$ plane for the incident light. The third contribution is the field of the radiation emitted by the dipole and back-scattered by the sample (image-dipole contribution). It can be  analytically calculated via a Fourier-decomposition of the dipole field into partial waves with various in-plane momenta $q$ and multiplying them by a respective Fresnel reflection coefficient \cite{AizpuruaOE08}:
\begin{equation}
   E_{\text{ima},z}= \frac{i d}{4\pi \varepsilon_{0}} \int_{0}^{\infty} r_{p}(q,\omega) (q^{3}/k_z)e^{i 2k_{z} z_{d}} dq,\label{EqEimaz}
   \end{equation}
\noindent where $k_{z}=(q_{0}^{2}-q^{2})^{1/2}$ is the vertical component of the momentum. 

By combining equations (\ref{Eqdz})-(\ref{EqEimaz}), we obtained a self-consistent expression for the dipole moment:

\begin{equation}
d(z_{d})= \frac{4\pi\varepsilon_0a^3 E_{0} F(z_{d})}{1-a^3 \int_{0}^{\infty} r_{p}(q,\omega)C(q,z_{d})dq}\label{EqdzSelfCons}
\end{equation}

\noindent where
\begin{equation}
F(z)=\left[e^{-i q_{0}z\cos\theta }+r_{p}(q_{\text{inc}},\omega)e^{i q_{0}z\cos\theta }\right]\sin{\theta} \label{EqF}
\end{equation}
\noindent is the illumination factor and
\begin{equation}
C(q,z)=(iq^{3}/k_z)e^{i2k_{z} z}
\end{equation}
\noindent is the dipole-sample coupling function. We note that in the near-field regime ($q\gg q_{0}$), $C(q,z_{d})\sim q^{2}e^{-2q z_{d}}$ leading to the maximum coupling at $q=1/(z_{d})$. Therefore the s-SNOM signal in many cases, including the one considered in this paper, is largely determined by the value of the Fresnel coefficient at this specific momentum (see Figs. 3d and 3f of the main text).

In the experiment, the dipole emission is captured by a remote detector. Importantly, the emitted radiation follows the same optical path as the one used for the dipole illumination, with the same polarization, angle of incidence, contribution from the sample reflection (dashed arrows in Fig. S2a). Therefore, the complex SNOM signal $\tilde{s}=s e^{i\phi}$ for a given dipole height is proportional to the dipole moment multiplied by the same illumination factor as used for the illumination path:
\begin{equation}
  \tilde{s}(z_d)\propto  d(z_{d})F(z_{d})\propto\frac{F^{2}(z_{d})}{1-a^3\int_{0}^{\infty} r_{p}(q,\omega)C(q,z_{d})dq}\label{EqSzd}
\end{equation}
\noindent up to an unimportant prefactor, which can be eliminated by normalization to a reference.

The s-SNOM/AFM instrument operates in the so-called tapping mode, where the tip oscillates vertically at a frequency $\Omega\ll\omega$ with the peak-to-peak amplitude $z_{\text{amp}}$ and the minimum distance $z_{\text{min}}$: $z_{d}(\varphi)=z_{\text{min}}+z_{\text{amp}}(1-\cos\varphi)$, where $\varphi=\Omega t$ is the oscillation phase. The oscillating optical signal is demodulated simultaneously at several higher harmonics $\Omega_{n}=n\Omega$. The s-SNOM output at a given harmonic $n$ can be calculated by the Fourier transform of $\tilde{s}(z_d)$:
\begin{equation}
  \tilde{s}_{n}=s_{n}e^{i\phi_{n}}=\frac{1}{2\pi}\int_{0}^{2\pi}e^{i n\varphi}\tilde{s}[z_d(\varphi)]d\varphi.
\end{equation}

The illumination factor given by Eq. (\ref{EqF}) depends on the sample properties and its influence on the s-SNOM maps in strongly anisotropic samples can be important, even after normalization to a reference. As it was shown in Ref. \cite{MesterNanophotonics22}, one can suppress this unwanted effect by calculating ratios between various harmonics (in this paper we use the value $\tilde{s_{r}}=\tilde{s_{3}}/\tilde{s_{2}})$. Indeed, in this double normalization, the factor $F^2(z_{d})$ in the Eq.(\ref{EqSzd}) is almost canceled because its variation with the dipole height is much slower than the one of the coupling function $C(q,z_{d})$ in the near-field limit $q\gg q_{0}$.

\begin{figure*}[htb]
\centerline{\includegraphics[width=12cm]{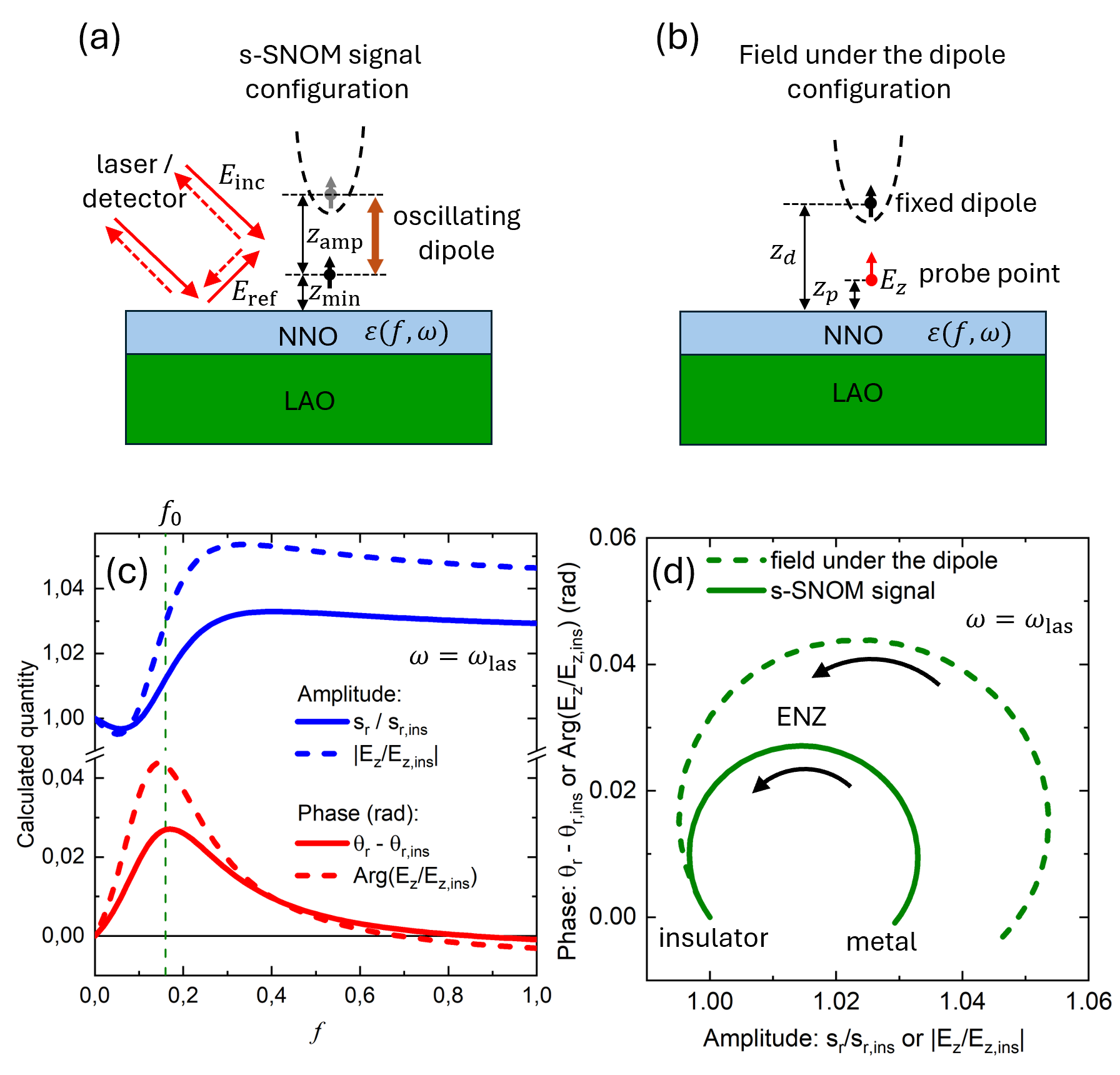}}

\begin{justify}
\textbf{Fig. S2.} (a) Complete model of the s-SNOM experiment involving external illumination (solid arrows), mechanical dipole oscillation and back-emission (dashed arrows). (b) Simplified model with a fixed dipole emitter and a probe field measured under the dipole. (c) the insulator-state normalized amplitude (blue) and phase (red) of the s-SNOM signal in the complete and simplified s-SNOM model (solid and dashed lines respectively). (d) Amplitude-phase correlations for the complete (solid line) and simplified (dashed lines) models, corresponding to panel (c).
\end{justify}

\label{Fig_SuppPointDipole}
\end{figure*}

A useful approximation for the actual s-SNOM signal consists of calculating the electric field under a fixed dipole (Fig. S2b) \cite{NikitinNPh16}. Here we treat the dipole as an emitter with a constant amplitude $d$ placed at a constant height $z_{d}$. We disregard the incident and reflected fields and consider only the direct field $E_{\text{dip},z}$ generated by the dipole and the field $E_{\text{ima},z}$ created by the image dipole. The direct field is equal to 
\begin{equation}
   E_{\text{dip},z}=\frac{d}{2\pi \varepsilon_{0}}\frac{(1-iq_{0}r)e^{iq_{0}r}}{r^{3}},
\end{equation}

\noindent where $r=z_{d}-z_{p}$. The image field is obtained by the mentioned above method of Fourier-decomposition:
\begin{equation}
   E_{\text{ima},z}= \frac{i d}{2\pi \varepsilon_{0}} \int_{0}^{\infty} r_{p}(q,\omega) C(q, (z_{d}+z_{p})/2) dq.
   \end{equation}
\noindent The total field is the sum of the two:
\begin{eqnarray}
   E_{z}&=&\frac{d}{2\pi \varepsilon_{0}}\left\{\frac{(1-iq_{0}r)e^{iq_{0}r}}{r^{3}}\right.\nonumber \\
   &+&\left.i \int_{0}^{\infty} r_{p}(q,\omega) C\left(q, \frac{z_{d}+z_{p}}{2}\right) dq\right\}.
\end{eqnarray}
\noindent Notably, only the second term depends on the sample properties. The first one contributes to unimportant constant background, but we keep it to enable direct comparison with the results of numeric FEM simulations.

In Fig. S2c, the results of the calculation described in the main text (Fig. 3e) using the complete model (solid lines) and simplified model (dashed lines) are compared. The corresponding amplitude-phase correlations are compared in Fig. S2d. One can see that the two calculations produce qualitatively matching results, which justifies using the simplified s-SNOM model for three-dimensional FEM simulations near the edge, thus saving significant computational resources.

\hspace{2pt}

\noindent \textbf{\large Note 4. Additional information for Fig. 2} 

\hspace{2pt}

\noindent Fig. S3 presents s-SNOM maps of amplitude and phase as well as the corresponding APCs for the sample described in Fig. 2a-c of the main text at two additional temperatures (95 K and 92 K). One can see that the plasmonic structures at these temperatures are qualitatively the same as those at 93 K, which are discussed in the main text.

Fig. S4 shows s-SNOM data at 13 K similar to Fig. 2g-i but on another device carrying the same current (10 mA). While the data are qualitatively similar, the phase peak in this device (FWHM = 115 nm) is somewhat narrower than that in device 1 shown in Fig. 2 (FWHM = 170 nm).

Fig. S5 compares the amplitude and phase maps and their profiles with the ones of the AFM topography. It shows clearly the absence of any topographic feature related to the metal-insulator boundaries.

\hspace{2pt}

\noindent \textbf{\large Note 5. Additional information for Figures 4 and 5} 

\hspace{2pt}

\noindent Fig. S6 presents the computed profiles of  amplitude and phase across an intermediate-width boundary ($w$ = 100 nm) for the case of low (a) and high (b) dipole position.

Fig. S7 shows the computed $xy$ maps of the electric field when a point dipole is situated in the middle of the metal-insulator boundary (as described in the main text), in the cases of sharp, intermediate and smooth boundaries. For comparison, the cases of uniform metal ($f$=1) and insulator ($f$=0) are shown.

\bibliography{ref}

\begin{figure*}[htb]
\centerline{\includegraphics[width=18cm]{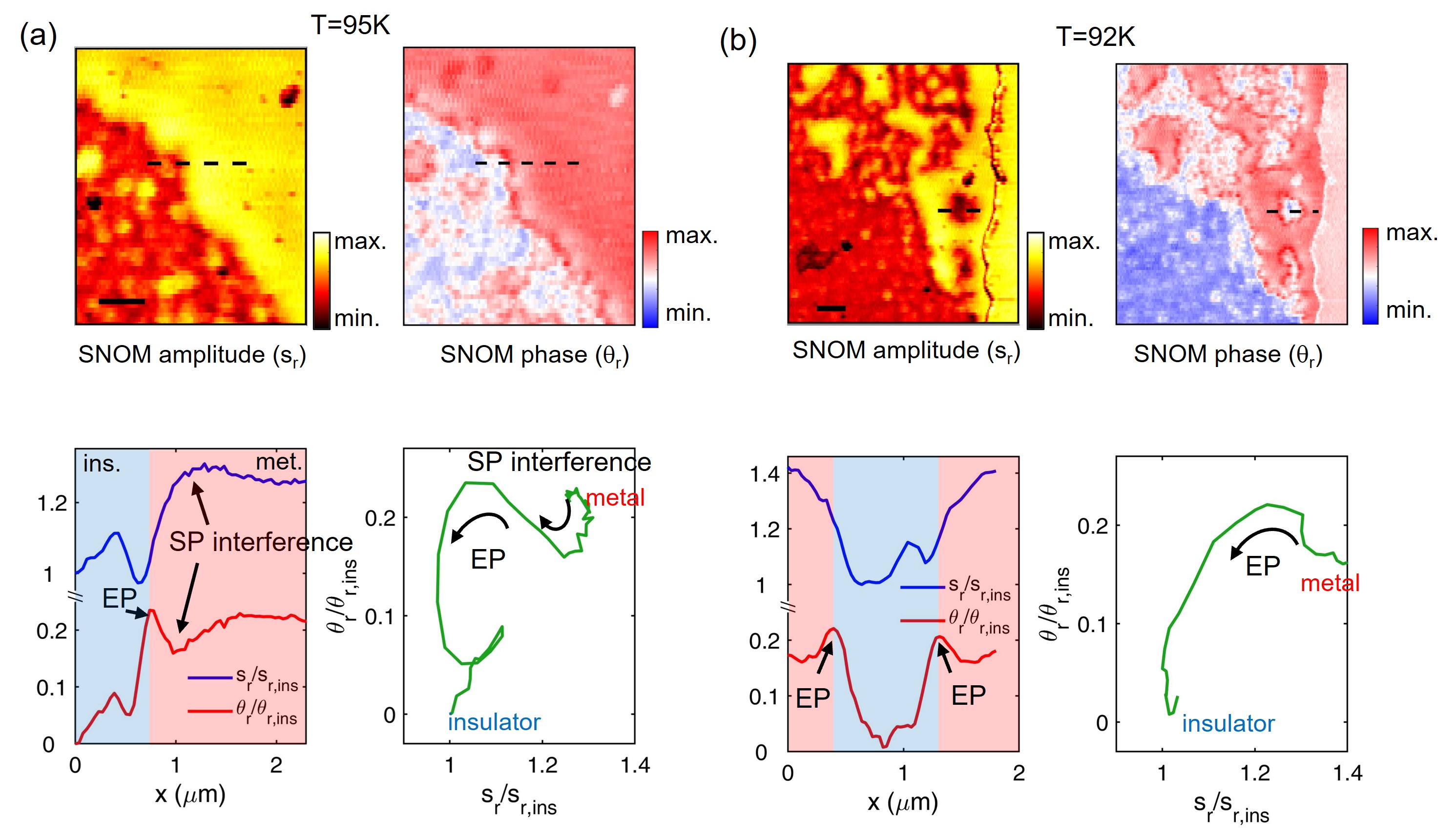}}
\begin{justify}
\textbf{Fig. S3.} s-SNOM images similar to the ones described in Fig. 2a,b,c of the main text, but for additional temperatures: 95~K (a) and 92~K (b). All scale bars are 1~$\mu$m.
\end{justify}

\label{SI_plasmons_twoT}
\end{figure*}

\begin{figure*}[htb]
\centerline{\includegraphics[width=6cm]{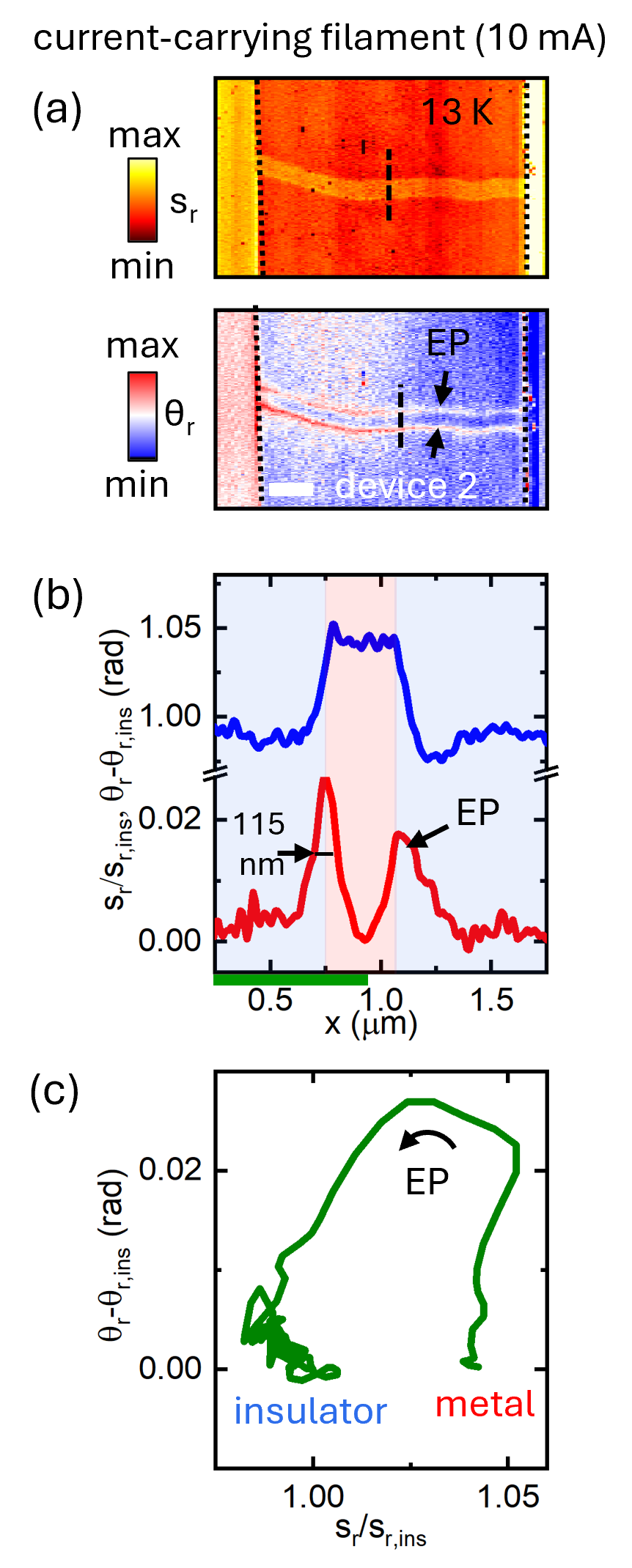}}
\begin{justify}
\textbf{Fig. S4.} s-SNOM maps (a), profiles (b) and amplitude-phase correlations (c) for the device 2 at 13 K carrying the current of 10 mA.  The scale bar is 2~$\mu$m. More explanations can be found in the caption to Fig. 2g-i of the main text.
\end{justify}

\label{Fig_SuppDevice2}
\end{figure*}

\begin{figure*}[htb]
\centerline{\includegraphics[width=10cm]{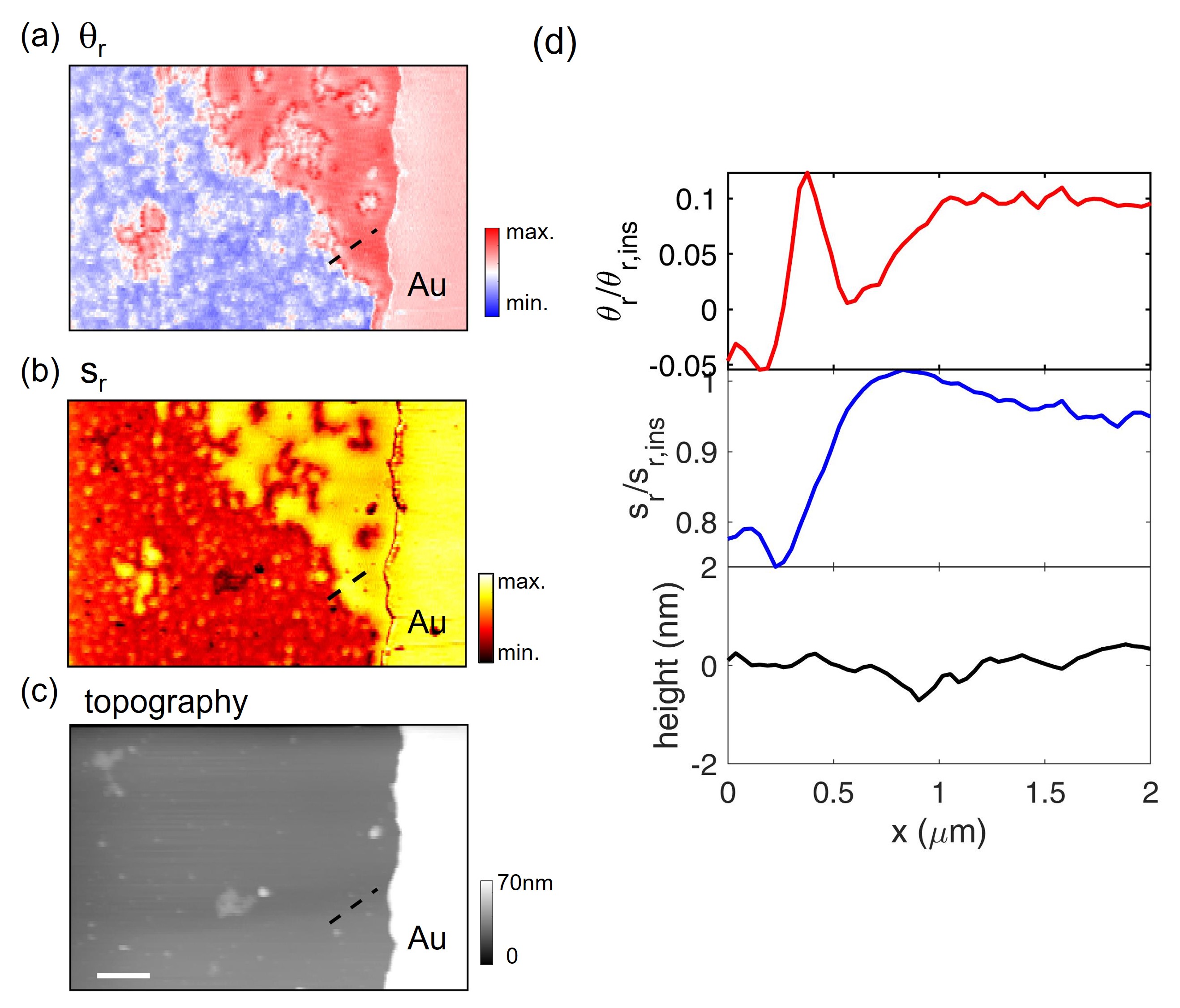}}
\begin{justify}
\textbf{Fig. S5.} Comparison between near-field amplitude(a), phase(b) and topography(c) signals at 93~K during cooling down. (d) The extracted profiles from the black dashed lines in (a-c). Scale bar, 2~$\mu$m. 
\end{justify}

\label{SI_topography_93K}
\end{figure*}

\begin{figure*}[htb]
\centerline{\includegraphics[width=10cm]{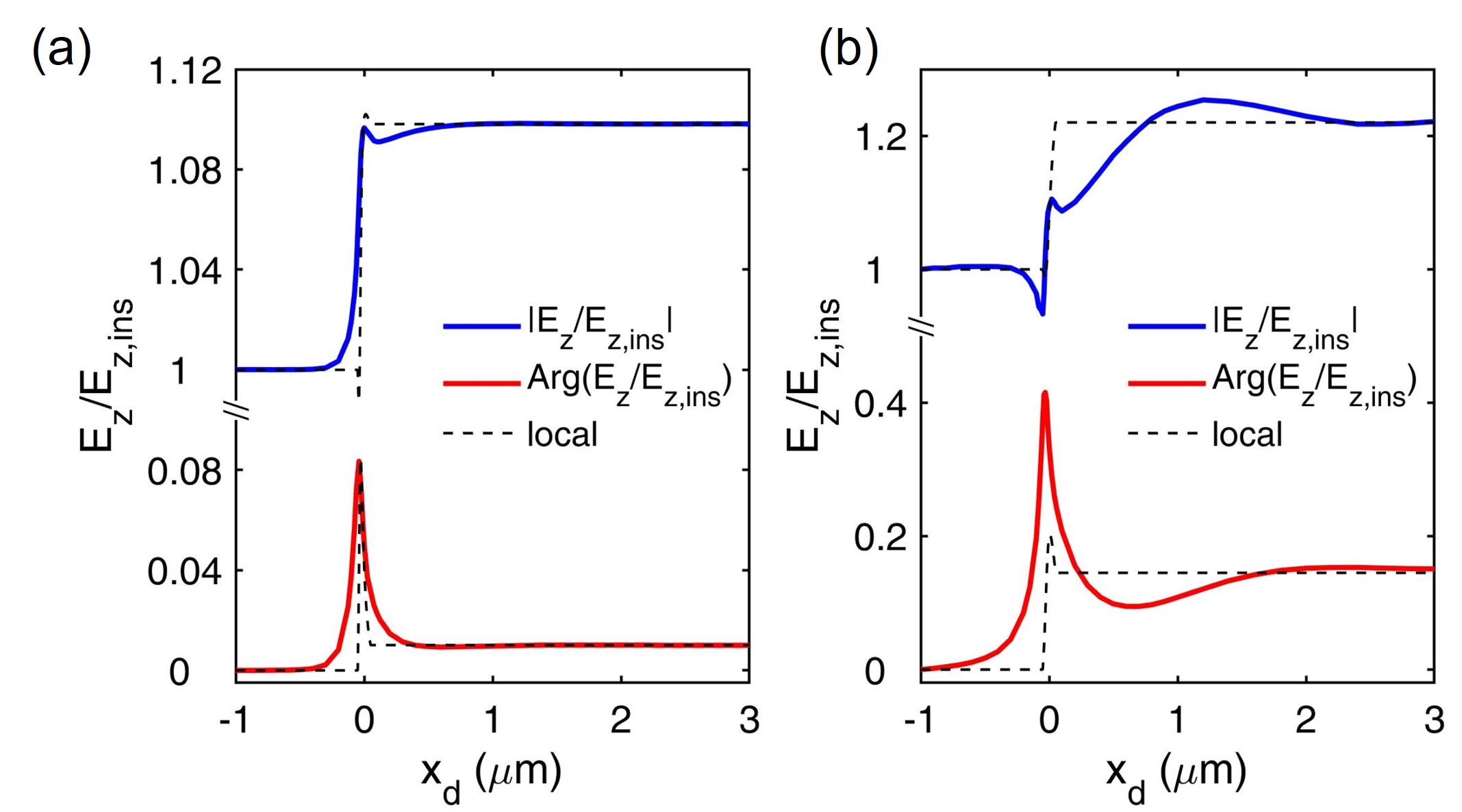}}

\begin{justify}
\textbf{Fig. S6.} Simulated near-field amplitude and phase profiles for the case of transition width $w$=100~nm, with dipole heights of $z_d$=0.2~$\mu$m (a) and 1.5~$\mu$m (b). The dashed-dotted line correspond to the \textit{gedanken} 'local' values defined in the main text. 
\end{justify}

\label{SI_w=100nm_withxd}
\end{figure*}

\begin{figure*}[htb]
\centerline{\includegraphics[width=16cm]{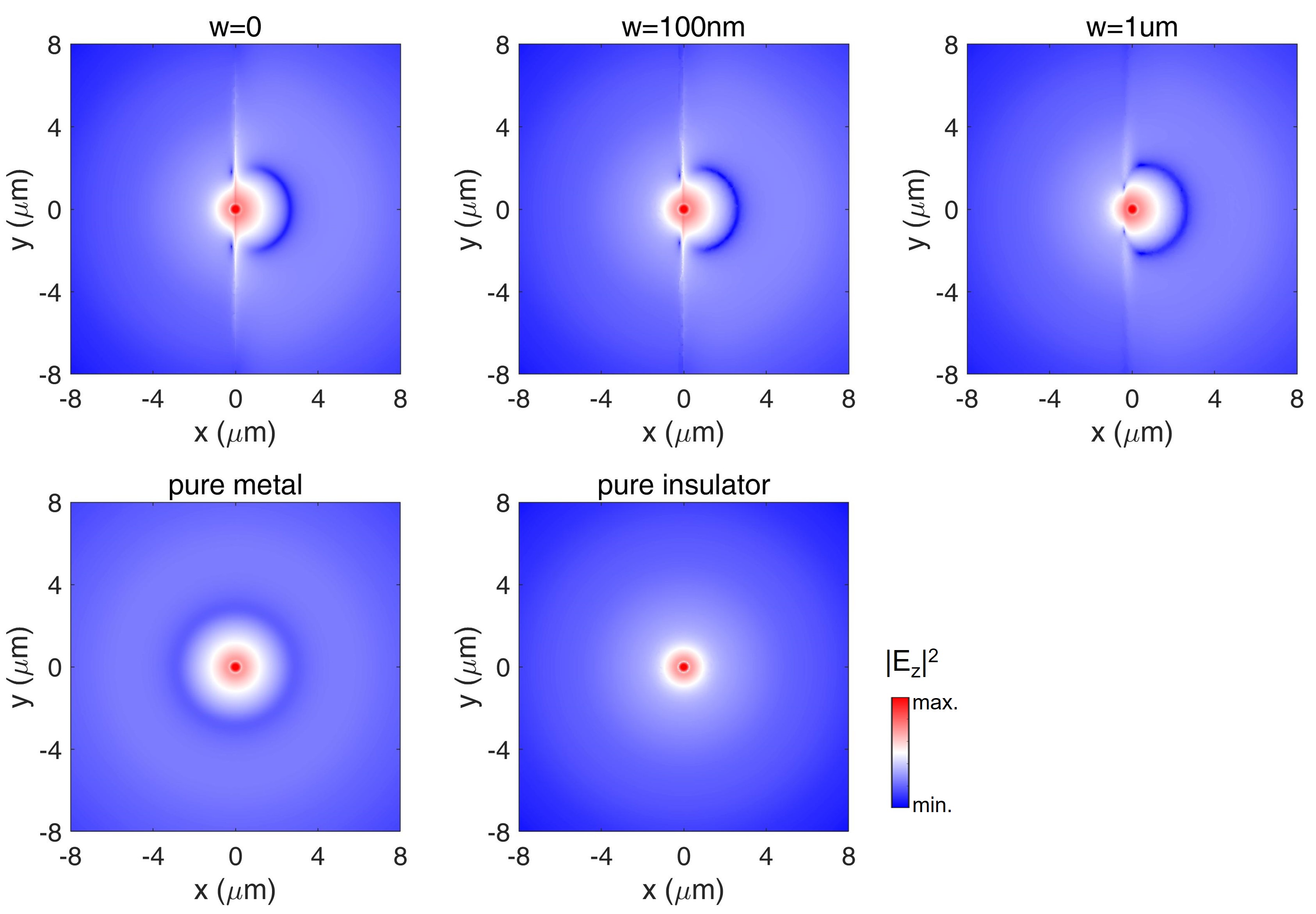}}
\begin{justify}
\textbf{Fig. S7.} Comparison of the electric field maps created by a dipole in the middle of the image (as described in the main text), in the cases of sharp, intermediate and smooth boundaries, as well as for a uniform metal and a uniform insulator.
\end{justify}

\label{SI_z0=0.2um_varied cases}
\end{figure*}